\documentstyle[12pt]{article}

\begin{document}
\title{\large {\bf A SIMPLE MODEL FOR THE NON-RETARDED DISPERSIVE FORCE BETWEEN
AN ELECTRICALLY POLARIZABLE ATOM AND A MAGNETICALLY POLARIZABLE ONE }}
\author{{\large C. Farina $^{\star}$, F.C. Santos$^{\dagger}$ and A.C. Tort$%
^{\ddagger}$} \\
Instituto de F\'{\i}sica - Universidade Federal do Rio de Janeiro - CP 68528
\\
Rio de Janeiro, RJ, Brasil - 21945-970.}
\maketitle
\begin{abstract}
It is well known that for the case of two neutral but electrically polarizable atoms the consideration or not of retardation effects on the dispersive van der Waals force between them leads essentially to different power laws for the forces; while the retarded force is proportional to $1/r^8$, where $r$ is the distance between the atoms, the non-retarded force is proportional to $1/r^7$. Here we consider the (repulsive) dispersive force between an electrically polarizable atom and a magnetically polarizable one and show  that, in contrast to the previous case, a quite unexpected result appears, namely: while the retarded  force is still proportional to $1/r^8$, the non-retarded force is proportional to $1/r^5$. We employ a semiclassical method based on the fluctuating dipole model for both atoms.
\end{abstract}

 \bigskip
\vfill
\noindent $^{\star }$ {e-mail: farina@i.f.ufrj.br}

\noindent $^\dagger$ {e-mail: filadelf@i.f.ufrj.br}

\noindent $^\ddagger$ {e-mail:tort@if.ufrj.br}

\pagebreak

\noindent 
In the context of thermodynamics of real gases, it was recognized by van der Waals \cite{VDW} in the second half of last century that attractive forces should exist between molecules (atoms) even when the molecules do not posses permanent electric dipole moments. Since these forces depend on the atomic (electric) polarizabilities, which in turn are closely related to the refractive index, they are referred to as dispersive van der Waals forces. When retardation effects of the electromagnetic field interaction are neglected, an approximation that can be made only for short distances, these forces are called non-retarded dispersive van der Waals forces, while the name retarded dispersive van der Waals forces is left for the cases where the retardation effects are not negligible. From experimental data, van der Waals tried to infer the form of the interaction energy between the atoms, but the correct form and explanation for the non-dispersive forces had to wait for the advent of quantum mechanics and only after the paper by London \cite{London30} in 1930 the precise origin of these forces were understood. The influence of retardation effects on the dispersive forces were obtained for the first time in 1948 by Casimir and Polder \cite{CasPol48}. Basically they showed that, while the non-retarded force is proportional to $1/r^7$, where $r$ is the distance between the two atoms, the retarded force behaves like $1/r^8$. An immediate consequence of dispersive forces between two atoms is the existence of (dispersive) forces also for macroscopic bodies (for a detailed discussion see for instance \cite{Langbein}). For the case of two rarefied macroscopic bodies,
 the resultant dispersive force can be obtained by a direct integration of the pairwise forces. On the other hand, for non-rarefied media the non-additivity effects must be taken into account  \cite{ Langbein,Milloni} (see also \cite{FST99} for a simple discussion). However, though the final numerical result is affected by these effects, the pairwise integration already gives the correct dependence on the geometry. For instance, the retarded force per unit area between two parallel semi-infinite slabs of polarizable material distant apart a distance $d$ is proportional to $1/d^4$, no matter the non-sdditivities are taken into account or not. Retardation effects are automatically taken into account when the interaction energy is calculated with the aid of the zero-point energy method, introduced in 1948 by Casimir \cite{Cas48}. In fact, the (standard electromagnetic) Casimir force between two macroscopic bodies can be considered as the retarded dispersive van der Waals force between them with the non-additivity effects included\footnote{The Casimir effect is more general than that, since it occurs whenever a relativistic field (not only the electromagnetic one) is constrained by some boundary conditions or is considered in a topologically non-trivial manifold \cite{Mostepanenko}}.

In this note, we shall investigate the non-retarded interaction between two non-similar atoms: one of them is only electrically polarizable and the other, only magnetically polarizable. We shall show that a quite unexpected result is obtained, namely: while the retarded interaction  is proportional to $1/r^7$ (retarded force proportional to $1/r^8$), the non-retarded interaction is proportional to $1/r^4$ (non-retarded force proportional to $1/r^5$),  in contrast to the $1/r^6$ dependence for the case of the interaction energy between two electrically polarizable atoms. Though here we discuss a simple model, this drastic change is fully justified in a QED calculation \cite{Nos}.

The retarded interaction energy between two atoms with both electrically and magnetically polarizations was calculated for the first time by Feinberg and Sucher \cite{Feinberg} and is given by (see also a detailed discussion in reference \cite{Boyer69}:

\begin{equation}
U(r)=\left[ -23\left( \alpha _1\alpha _2+\beta _1\beta _2\right) +7\left(
\alpha _1\beta _2+\alpha _2\beta _1\right) \right] \frac{\hbar c}{4\pi r^7}%
\,,  \label{boyer}
\end{equation}
where in Eq.(\ref{boyer} ) $\alpha _i$ is the (static) polarizability
and $\beta _i$ is the (static) magnetic polarizability, but, as far as the authors are aware of, no one in literature had considered the non-retarded case until very recently \cite{Nos}, where the authors show, with the aid of perturbative quantum electrodynamics techniques, that the non-retarded dispersion van der Waals interaction potential for the case at hand obeys a $1/r^4$ law and not the $1/r^6$ law as one would naively think. It is the purpose of this note to obtain this rather surprising result in a much simpler way, which we think can be understood by any undergraduate student who is familiar with introductory electromagnetism and ordinary quantum mechanics.

Our procedure will be based on the dipole fluctuating method as discussed for instantce in references \cite
{Langbein,Milloni,FST99}. To this effect we model the electric polarizable molecule by an
electric charge $e$ of mass $m_e$ which oscillates harmonically along a fixed, but arbitrary direction defined by the unit vector ${\bf \hat{u}}_e$, so
that the atom is effectively replaced by an electric dipole ${\bf p}%
(t)=ex_e(t{\bf )\hat{u}}_e$, where $x_e$ measures the position along the
oscillation axis taking the origin at the equilibrium position. The magnetically polarizable atom will be replaced by
a magnetic charge $g$ of mass $m_g$ which, similarly to its electric
counterpart, also oscillates harmonically along another fixed, but arbitrary direction
defined by the unit vector ${\bf \hat{u}}_g$. Effectively we have a magnetic
dipole ${\bf m}(t)=g\,x_g(t)\,{\bf \hat{u}}_g$ playing the role of the
magnetically polarizable atom. Each dipole is the source of electric and
magnetic fields that pervade all space. Each dipole also probes the electric
and magnetic fields created by the other. 

The electric and magnetic fields at a point of the space identified by the
position vector ${\bf r}$ created by an electric dipole  are given by \cite{Jackson}: 
\begin{eqnarray}
{\bf E}_q({\bf r},t) &=&\left[
3\left( {\bf p}(t^{*})\cdot \hat{r}\right) \hat{r}-{\bf p}(t^{*})\right]
\frac 1{r^3}+\left[ 3\left( {\bf \dot{p}}(t^{*})\cdot \hat{r}\right) \hat{r}-{\bf \dot{%
p}}(t^{*})\right] \frac 1{cr^2}\nonumber \\
&&-\left[ {\bf \ddot{p}}(t^{*})-\left( {\bf \ddot{p}}%
(t^{*})\cdot {\bf \hat{r}}\right) {\bf \hat{r}}\right] \frac 1{c^2r}\,,
\end{eqnarray}
and 
\begin{equation}
{\bf B}_q({\bf r},t)=\left[ \frac 1{cr^2}{\bf \dot{p}}(t^{*})+\frac 1{c^2r}%
{\bf \ddot{p}}(t^{*})\right] \times {\bf \hat{r}}\,.
\end{equation}
where $r$ is the distance from the dipole to the point of observation and $%
t^{*}=t-r/c$  is the retarded time. The fields generated by a
magnetic dipole can be obtained by means of the substitutions ${\bf p}\to 
{\bf m}$, ${\bf E}\to {\bf B}$ and ${\bf B}\to -{\bf E}$ \cite{Jackson}. The
Lorentz force acting on the electric dipole whose position is identified by
the vector ${\bf r}_e$ at a given instant of time $t$ is 
\begin{equation}
{\bf F}_e({\bf r}_e,t)=e{\bf E}_g({\bf r}_e,t^{*})+\frac ec{\bf v}%
_e(t)\times {\bf B}_g({\bf r}_e,t^{*})\,.
\end{equation}
In the same way, the Lorentz force acting on the magnetic dipole will be
given by:
\begin{equation}
{\bf F}_g({\bf r}_g,t)=g{\bf B}_e({\bf r}_g,t^{*})-\frac gc{\bf v}%
_g(t)\times {\bf E}_e({\bf r}_g,t^{*}).\,
\end{equation}
If we project these forces along the directions defined by the (fixed) unit
vectors ${\bf \hat{u}}_e$ and ${\bf \hat{u}}_g$, neglect retardament and
radiation reaction, and take into account the forces that bind the
oscillating electric and magnetic charges to their respective centers of
force by simulating them through Hookean forces, we end up with a coupled
system of linear differential equations which reads 
\begin{eqnarray}
\ddot{x}_e(t)+\omega _e^2x_e(t)=\frac e{m_e}{\bf E}_g({\bf r}_e,t)\cdot {\bf 
\hat{u}}_e+\frac e{m_ec}{\bf v}_e(t)\times {\bf B}_g({\bf r}_e,t)\cdot {\bf 
\hat{u}}_e\,,
\end{eqnarray}
\begin{eqnarray}
\ddot{x}_g(t)+\omega _g^2x_g(t)=\frac g{m_g}{\bf B}_e({\bf r}_g,t)\cdot {\bf 
\hat{u}}_g+\frac g{m_gc}{\bf v}_g(t)\times {\bf E}_e({\bf r}_g,t)\cdot {\bf 
\hat{u}}_g\,.
\end{eqnarray}
where $\omega _e$ and $\omega _g$ are, respectively, the natural frequencies
of the electric and magnetic dipoles. If the dipoles are not very far from
each other ($r\ll \,137a_0,$ where $a_0$ is Bohr's radius, see for instance \cite{FST99}),  
the dominant term leads to:
\begin{equation}
\ddot{x}_e(t)+\omega _e^2x_e(t)=F(r)\,\,\dot{x}_g(t)  \label{eqm1}
\end{equation}
\thinspace \thinspace \thinspace \thinspace \thinspace \thinspace \thinspace
\thinspace \thinspace \thinspace \thinspace \thinspace \thinspace 
where $%
F(r):=+\ \frac{eg}{m_ecr^2}(\hat{u}_g\times \hat{r}_{eg})\cdot \hat{u}_e$,
since $\hat{r}_{ge}=-\hat{r}_{eg}$, and 
\begin{equation}
\ddot{x}_g(t)+\omega _g^2x_g(t)=H(r)\,\,\dot{x}_e(t)  \label{eqm2}
\end{equation}
where $H(r):=+\ \frac{eg}{m_gcr^2}(\hat{u}_e\times \hat{r}_{eg})\cdot \hat{u}%
_g$. Equations (\ref{eqm1}) and (\ref{eqm2}) form system of coupled
differential equations that can be solved in the usual way, {\it i.e.,} we
try a solution of the form $x_i(t)=C_i\exp (-i\omega t)$, where $C_i\,$ is a
complex constant$\,$with $i=e,g$, thereby obtaining the following algebraic
equation: 
\begin{equation}
\Omega ^2-\,\left( \omega _e^2+\omega _g^2-F(r)H(r)\right) \Omega +\;\omega
_e^2\omega _g^2=0.  \label{algebricequation}
\end{equation}
where $\Omega :=\omega ^2.$ If we set $F(r)H(r)=0$ we can easiliy verify
that the roots of the above equation are real and given by $\Omega _1=\omega
_e^2$ and $\Omega _2=\omega _g^2$ as expected. The roots of Eq.(\ref
{algebricequation} ) satisfy the simple relations 
\begin{equation}
\Omega _1+\Omega _2=\omega _e^2+\omega _g^2-F(r)H(r)  \label{propri1}
\end{equation}
and 
\begin{equation}
\Omega _1\,\Omega _2=\omega _e^2\,\omega _g^2  \label{propri2}
\end{equation}
Let us introduce the auxiliary quantity 
\begin{equation}
S:=\left( \omega _1+\omega _2\right) ^2-\left( \omega _e+\omega _g\right) ^2,
\end{equation}
where $\omega _1$ and $\omega _2$ are two real positive roots of the
original algebraic equation of the fourth degree for $\omega $. Using the
simple properties given by Eqs.(\ref{propri1}) and (\ref{propri2}) we find 
\begin{equation}
S=-F(r)H(r)
\end{equation}
Hence from and we can write 
\begin{equation}
\omega _1+\omega _2=\left( \omega _e+\omega _g\right) \sqrt{1+\frac S{\left(
\omega _e+\omega _g\right) ^2}}
\end{equation}
Upon quantization and assuming that $\left| S\right| \ll \left( \omega
_e+\omega _g\right) ^2$ the ground state energy will be given by

\begin{eqnarray}\label{energy}
E_0 &=&\frac \hbar 2\left( \omega _1+\omega _2\right) \approx \frac \hbar
2\left( \omega _e+\omega _g\right) +\frac{\hbar \,\,e^2g^2}{4\,\left( \omega
_1+\omega _2\right) \,m_{e\,}m_{g\,}c^2}\frac{Q_{eg}^2}{r^4}\nonumber  
\\
&=&\frac \hbar 2\left( \omega _e+\omega _g\right) +\frac \hbar 4\left( \frac{%
\omega _{e\,}\omega _g^{}}{c^2}\right) \left( \frac{\omega _{e\,}\omega _g^{}%
}{\omega _e+\omega _g}\right) \,\,\,\,\,\,\,\frac{\alpha \,\beta \,Q_{eg}^2}{%
r^4}\,
\end{eqnarray}
where we have rewritten the last term, which is the interaction potential,
so that the electric and magnetic polarizabilities $\alpha =\frac{e^2}{%
m_e\;\omega _e^2}$ and $\beta =\frac {g^2}{m_g\;\omega ^2}$ , can be identified
. The quantity \break $Q_{eg}:=({\bf \hat{u}}_g\times {\bf \hat{r}}_{eg})\cdot {\bf 
\hat{u}}_e$ is a spatial orientation factor to be properly averaged. The
force between the electrically polarizable atom and the magnetically
polarizable one is repulsive as expected, but it varies with the inverse
fifith power of the separation: 
\begin{equation}
{\bf F(}r)=\hbar \,\left( \frac{\omega _{e\,}\omega _g}{c^2}\right) \left( 
\frac{\omega _{e\,}\omega _g}{\omega _e+\omega _g}\right) \,\,\,\,\,\,\,%
\frac{\alpha \,\beta \,Q_{eg}^2}{r^5}\,{\bf r}_{eg}  \label{force1}
\end{equation}

The $1/r^4$ behavior of (\ref{energy}) (or the $1/r^5$ behavior of (\ref
{force1}) can be traced back to the near field approximation of the
electromagnetic fields involved, which contribute to the problem with terms
varying with $1/r^2$ and to the fact that terms stemming from the vector
products in the equations of motion are zero. Take, for instance, the magnetically polarizable atom. There are two forces acting on it: the first one is due to the magnetic field created by the fluctuating electri dipole of the other atom, which does not have the typical  $1/r^3$ behaviour of a static zone field; the second one is due to the cross product between the magnetic charge velocity (${\vec v}_g$) and the electric field generated by the electric fluctuating dipole (${\vec E}_e$). However, though ${\vec E}_e$ exhibits the static zone term (proportional to $1/r^3$), this term disappears from the calculation when projected into the fixed (but arbitrary) direction of motion of the magnetic charge $g$. An analogous explanation holds for the electric polarizable atom. We conclude, then, that this is a very peculiar result whose origin lies in the fact that one of the atoms is only electrically polarizable and the other, only magnetically polarizable. Had we started with two magnetically polarizable atoms, we would have obtained a non-dispersive interaction energy proportional to $1/r^6$.

\noindent {\bf Acknowledgements:} One of us (C.F.) wishes to acknowledge the
partial financial support of CNPq (the National Research Council of Brazil).

\end{document}